\documentclass[traditabstract]{aa}
\usepackage{graphicx}
\usepackage{natbib,twoopt,lscape}
\usepackage[breaklinks=true]{hyperref} 
 \bibpunct{(}{)}{;}{a}{}{,}    
 \newcommandtwoopt{\citeads}[3][][]{\href{http://adsabs.harvard.edu/abs/#3}%
                                        {\citealp[#1][#2]{#3}}}
 \newcommandtwoopt{\citepads}[3][][]{\href{http://adsabs.harvard.edu/abs/#3}%
                                        {\citep[#1][#2]{#3}}}
 \newcommandtwoopt{\citetads}[3][][]{\href{http://adsabs.harvard.edu/abs/#3}%
                                        {\citet[#1][#2]{#3}}}
 \newcommandtwoopt{\citeyearads}[3][][]%
   {\href{http://adsabs.harvard.edu/abs/#3}{\citeyear[#1][#2]{#3}}}
\bibpunct{(}{)}{;}{a}{}{,}
\usepackage{latexsym}
\usepackage{longtable}
\usepackage{amsmath}
\usepackage{amssymb}
\usepackage{amstext}
\usepackage{color}
\usepackage{psfrag}
\usepackage{txfonts}

\def\cm3{cm$^{-3}$}

\def\12{$^{12}$CO}

\def\nodata{...}

\begin{document}

\title{Feature-tailored spectroscopic analysis of the SNR Puppis~A in X-rays}


   \author{G. J. M. Luna\inst{1},  M. J. S. Smith\inst{2,3}, G. Dubner\inst{1}, E. Giacani\inst{1,4} \and G. Castelletti\inst{1}}

   \institute{Instituto de Astronom\'ia y F\'isica del Espacio (IAFE), CC 67 - Suc. 28 (C1428ZAA)  CABA -- Argentina.\\
              \email{gjmluna@iafe.uba.ar}
         \and
         XMM-Newton Science Operations Centre, ESAC, Villafranca del Castillo, 28080 Madrid, Spain
         \and
         Telespazio Vega U.K. S.L.         
         \and
    	 FADU, University of Buenos Aires, 1428 Buenos Aires, Argentina
         }

   \date{}

 
  \abstract
{We introduce a distinct method to perform spatially-resolved spectral analysis of astronomical sources with highly structured X-ray emission.
The  method measures the surface brightness of neighbouring pixels to adaptively size and shape each region, thus the spectra from the bright and faint filamentary structures evident in the broadband images can be extracted.  As a test case, we present the spectral analysis of the complete X-ray emitting plasma in the supernova remnant Puppis~A observed with XMM-{\it Newton} and {\em Chandra}. Given the angular size of Puppis~A, many pointings with different observational configurations have to be combined, presenting a challenge to any method of spatially-resolved spectroscopy. From the fit of a plane-parallel shocked plasma model we find that temperature, absorption column, ionization time scale, emission measure and elemental abundances of O, Ne, Mg, Si, S and Fe, are smoothly distributed in the remnant. Some regions with overabundances of O--Ne--Mg, previously characterized as ejecta material, were automatically selected by our method, proving the excellent response of the technique. This method is an advantageous tool for the exploitation of archival X-ray data.}
   
   \authorrunning{Luna et al.}
\titlerunning{X-ray spectroscopy of Puppis~A in X-rays}

   \maketitle
%

\section{Introduction}

Extended X-ray emitting plasmas are found in different astronomical scenarios such as novae and supernova remnants (SNR), galaxies and clusters of galaxies, as well as in outflows from accreting objects \citepads[e.g.]{2009ApJ...707.1168L,2015ApJ...801...92T,2012ApJ...746..130H,2013Sci...341.1365S,2008A&A...478..797G}. X-ray spectra from these extended regions are the ideal tool to investigate the nature of the emitting plasma, its ionization state, temperature, density and chemical composition. In particular, the X-ray radiation associated with SNRs can be thermal in nature, originating in a plasma heated by the passage of fast SNR shocks, and also of non-thermal, synchrotron origin associated with electron cosmic-ray acceleration.  The interaction of the shocks with the surrounding circumstellar and interstellar media plus the intrinsic properties of the explosion imprints the complex structures often observed in the X-ray images of SNR \citepads[e.g. G320.4$-$1.2;][]{2002ApJ...569..878G}. Thus, the study of these structures provides information about the progenitor as well as the different stages of the remnant evolution, from the onset of the explosion until the merging with the interstellar medium (ISM). 

Spectral analysis of these structures relies on the extraction of data from suitably defined spatial regions. The construction of these regions is usually based on some prior knowledge of local physical conditions (e.g. brightness, chemical composition and velocity) and in general consists of polygons. In the case of very structured X-ray emission, as  observed in some galactic SNRs, regular grids have been used to study the spatial distribution of plasma properties \citepads[see e.g.][]{2004A&A...427..199C,2008ApJ...676..378H, 2012ApJ...746..130H}. The accuracy of the spectral fits, however, is frequently limited by the surface brightness of the respective region, thus dramatically limiting the study of the faintest portions of a remnant. Recently, \citetads{2015MNRAS.453.3954L} presented a method to create small tessellated regions, whose shape is determined by an adaptive mesh that adds pixels counts until a certain threshold is reached.

The SNR Puppis~A is one of the brightest X-ray remnants in our Galaxy. It was detected for the first time more than 30 years ago using the {\em Einstein} Observatory \citep{petre1982}, and later thoroughly investigated using {\em ROSAT} \citepads{1993AdSpR..13...45A}, {\em Chandra} \citepads{2005ApJ...635..355H}, {\em Suzaku} \citepads{2008ApJ...676..378H}, and XMM-{\em Newton} \citepads{2006A&A...457L..33H,2010ApJ...714.1725K,2012ApJ...756...49K} satellites. However, because of its large angular size (about 50 arcmin), it was not until very recently that Puppis~A was observed in great detail in its entirety in the X-ray regime \citepads[][hereafter D13]{2013A&A...555A...9D}. The X-ray emission detected in Puppis~A is completely thermal in origin in spite of the presence of the central compact object RX~J0822$-$4300. In X-rays, the whole remnant appears very structured, being composed of multiple short filaments suggesting that it is evolving in a complex ambient medium. Two bright knots, the bright eastern knot (BEK) and the bright northern knot \citep{petre1982} are clear evidence of the interaction of the SNR shock with ambient clouds in a relatively late phase of evolution \citepads{2005ApJ...635..355H,2010ApJ...714.1725K,2012ApJ...756...49K}. Other isolated X-ray features rich in O, Ne and Mg have been identified as SN ejecta \citepads{2008ApJ...676..378H,2008ApJ...678..297K,2010ApJ...714.1725K}.

Knowledge of the spatial distribution of plasma temperature, ionization time scales and elemental abundances from X-ray observations contribute to the reconstruction of the history of the SNR. Using observations of Puppis~A obtained with XMM-{\em~Newton} and {\em Chandra}, \citetads{2010ApJ...714.1725K} studied the radial distribution of model parameters in pie-shaped regions centered on the inferred expansion center of the SNR. \citetads{2008ApJ...676..378H} used {\em Suzaku}/XIS observations of 5 fields throughout a portion toward the North-East of Puppis~A and studied the spatial distribution of model parameters in a grid of 259 square and rectangular regions with sizes of 100$^{\prime\prime}$--200$^{\prime\prime}$. Although this was a pioneering work on X-ray spectral analysis of Puppis~A, studying a substantial portion of the remnant, the authors were not able to obtain statistically acceptable fits because of the uncertainties in the spectral responses and low number of counts at small spatial scales. Our study focuses, for the first time, on the filamentary features of the remnant and derives statistically significant model parameters.


In this paper we present a distinct method for selecting the regions from which the spectra are extracted which is based on an adaptively-sized selection of regions
driven by the remnant's surface brightness. Spectral modeling thus yields parameter estimates which closely trace the brightness distribution of the remnant as observed in wide band X-ray images. 
In Section \ref{sec:obs} we describe the technique used to extract the spectra over the SNR. Our main results are presented in Section \ref{sec:results} while conclusions are summarized in Section \ref{sec:concl}.


\section{Data analysis strategy}
\label{sec:obs}

Our aim is to perform a spatially resolved spectral analysis of the SNR Puppis~A in order to study the distribution of plasma temperature, ionization time scale, absorbing column and elemental abundances. 
To this end, we used the combined data of 59 new and archival XMM-{\it Newton} and {\em Chandra} observations as described in D13. Per observation, the respective counts images were cleaned, divided by exposure time and corrected for vignetting. They were subsequently divided by the effective area (averaged over energy using the source spectrum as weight)
and finally combined to produce the image of the complete remnant in the 0.3 - 8.0 keV energy band (see Fig. 2 in D13).

This image was then re-sampled to lower resolution with square spatial bins of n arcsec in width. Starting from
the brightest bin, polygonal regions were created by iteratively adding the adjacent pixel with maximum brightness, repeating until the total number of counts within the respective polygon reached a predefined minimum threshold, $T$. Any remaining inter-polygonal areas with fewer than $T$ counts were combined with the adjacent polygon closest in average brightness. This strategy results in a tessellated mapping of the complete remnant in a set of contiguous polygons which trace the surface brightness morphology and which are, moreover, adaptively sized so as to contain sufficient counts to allow spectral analysis. 

Our selection method differs in principle from that presented by \citeads{2015MNRAS.453.3954L} (hereafter L15) in the criterion used for grouping neighboring pixels to form a region. L15 select pixels of similar counts, whereas we choose to group pixels of similar brightness. In the case of uniform coverage and region sizes which are small with respect to the telescope field-of-view, these two methods should give similar results. However, the data presented here are obtained from a highly non-uniform source coverage, as many portions of the remnant were observed on multiple occasions, while other areas were observed only once (see Table 1 in D13). Moreover, some of the resulting regions are sufficiently large to be subject to significant differential telescope vignetting. Hence, our method to construct the regions from where spectra will be extracted is preferable as it limits observational and instrumental biases and favors a mapping based on a source physical property. Thus our technique becomes an excellent tool for exploiting archival data adquired with different telescopes and in different epochs.

\begin{figure*}
\includegraphics[scale=1.0]{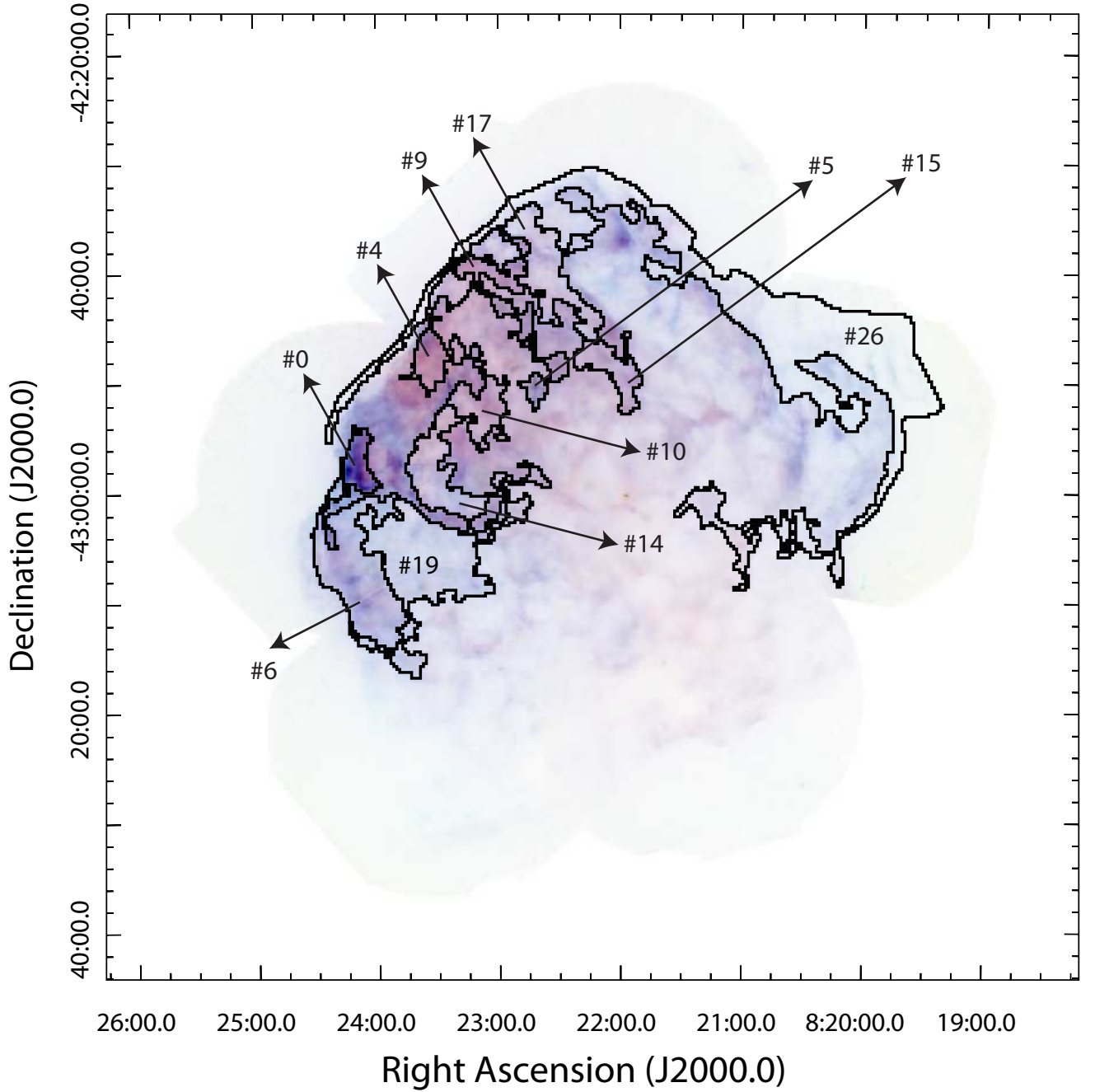}
\caption{Sample regions from the 35-regions case whose sizes and shapes were obtained with our ``feature-tailored'' technique. This figure displays the regions (black contours) and its corresponding identification number over the 0.3-8.0 keV image.}
\label{fig:panel1}%
\end{figure*}

\begin{figure*}
\includegraphics[scale=1.0]{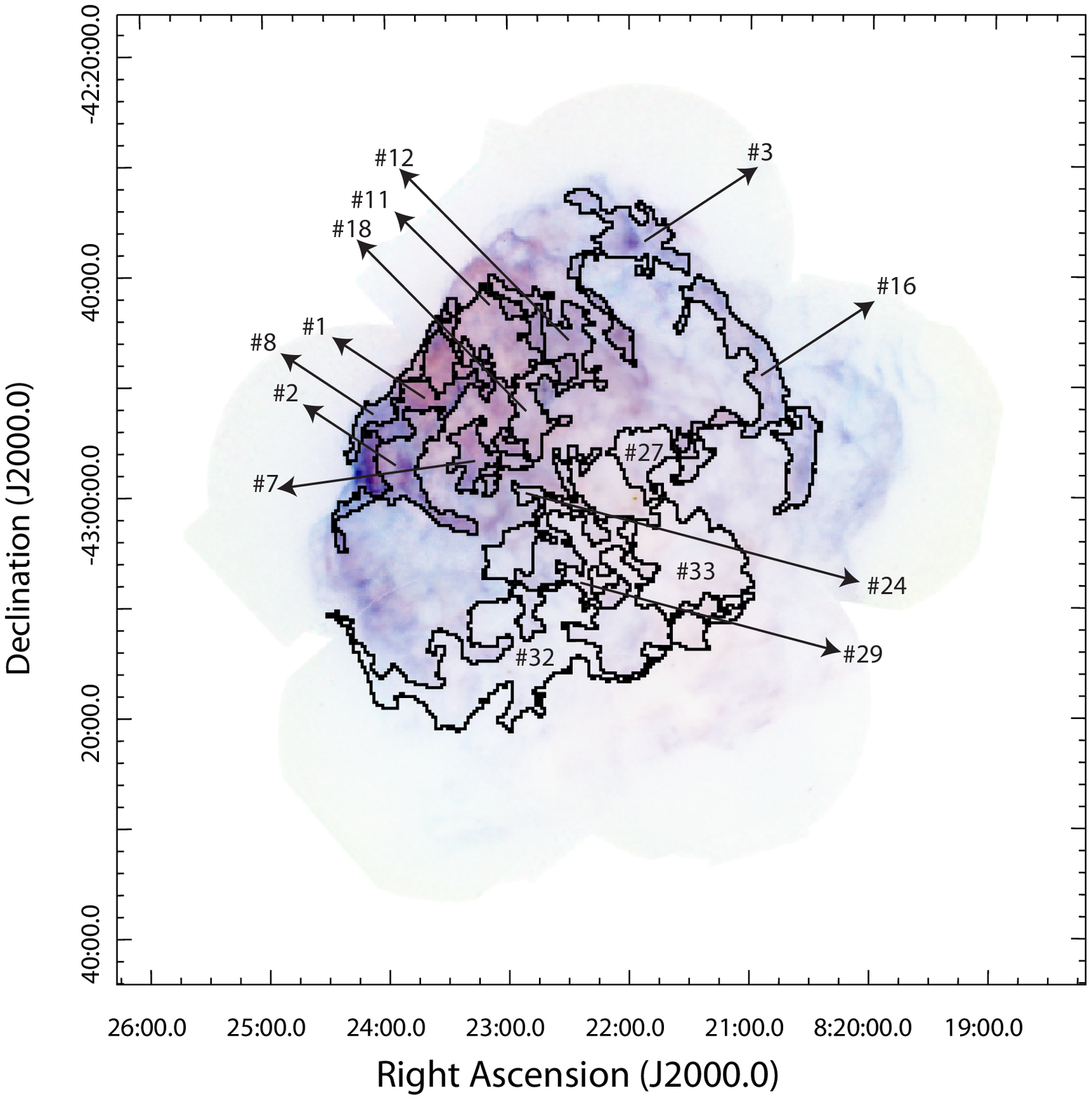}
\caption{Same as Fig. \ref{fig:panel1}}
\label{fig:panel2}%
\end{figure*}

\begin{figure*}
\includegraphics[scale=1.0]{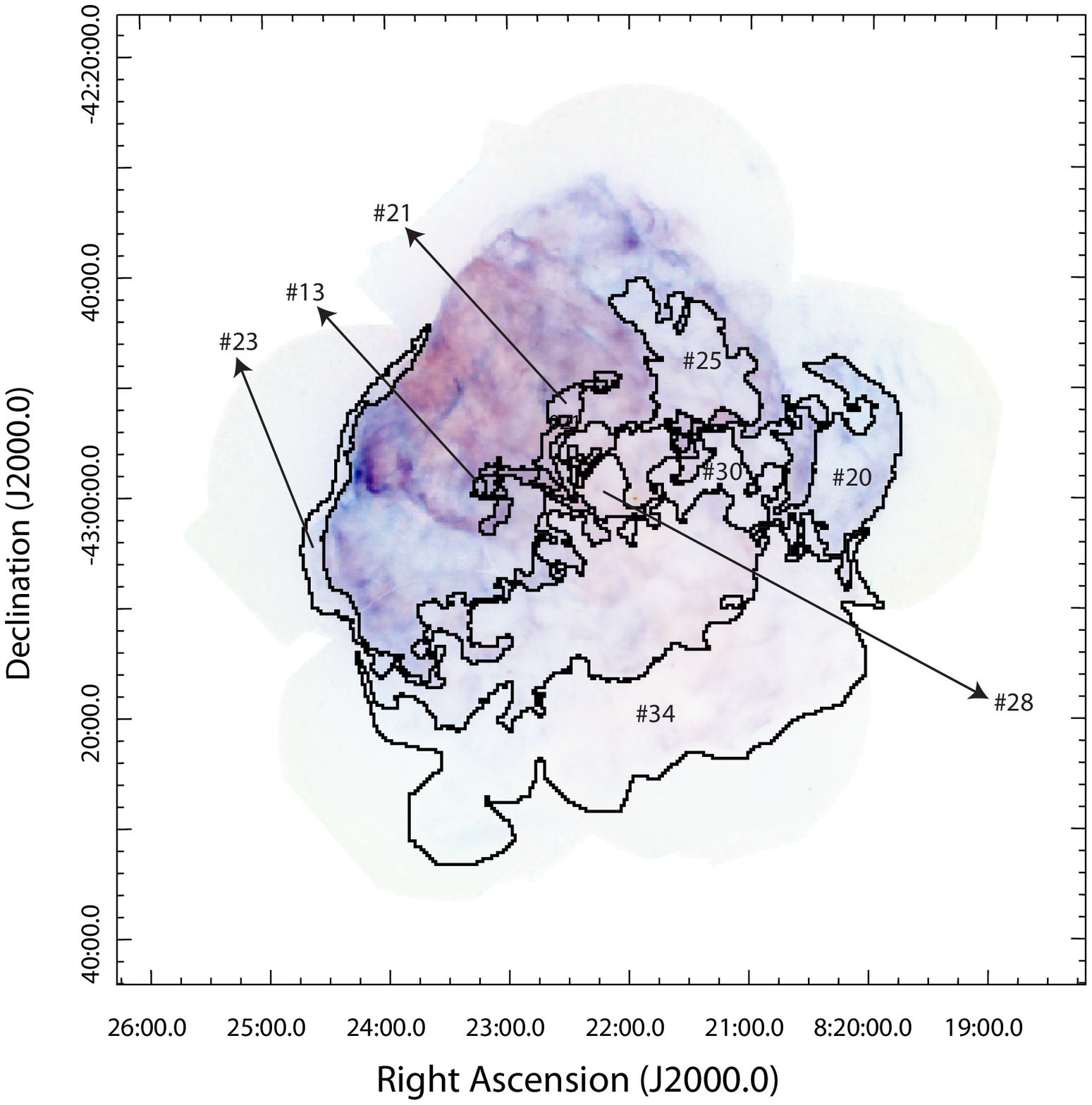}
\caption{Same as Fig. \ref{fig:panel1}}
\label{fig:panel3}%
\end{figure*}

A circular region of 1 arcmin radius around the central compact object was excluded from the data prior to defining the polygons. The reason for the above mentioned initial spatial re-sampling is to avoid polygon definitions which are overly complex with respect to the telescope and instrument spatial resolution. $T$ was determined from counts in the 0.3--8 keV band, and limited to mono pixel events for the {\em EPIC} instruments in view of spectral quality; standard pattern grades were used for {\em ACIS}. {\em EPIC-pn} spectra were corrected for out-of-time events. 
Montage sets of regions for three separate values of threshold $T$ (and slightly varying bin size $n$) were created. Respective values of $T$ of 2$\times$10$^{6}$, 1$\times$10$^{6}$, and 2$\times$10$^{5}$ counts and $n$ of 15, 20, and 15 arcsec resulted in 35, 96 and 419 individual regions (the respective values of n were optimised through trial-and-error, mainly to minimise the number of disjoint regions with few counts). Examples of resulting regions, which tend to closely follow the arched filamentary morphology of the remnant, are shown in Figs.~\ref{fig:panel1}, \ref{fig:panel2}, \ref{fig:panel3}.

For a given mesh, and per region, we extracted spectra from all overlapping observations and created the respective response files for each such spectrum. Background spectra were obtained from source free regions within the respective observations, or, in the case no such regions were available, from background templates scaled for exposure time. 

The brightest X-ray regions in Puppis~A (\#0 and \#1 in the 35-regions case) are prone to suffer from pile-up. In these cases, we removed the $EPIC-pn$, Full Window mode spectra from the fits. We rebinned the spectra to a minimum of 20 counts per bin and then fit, per region, a model consisting of an absorbed, plane-parallel shocked plasma with variable abundances \citepads[\texttt{vpshock};][and references therein]{2001ApJ...548..820B}. We used version 2.0 of \texttt{vpshock}, which utilizes atomic data from ATOMDB (Foster et al. 2012). Abundances of elements with prominent lines in the 0.3-8.0 keV band such as O, Ne, Mg, Si, S, and Fe were allowed to vary during the fit while C, N, Ar, Ca and Ni were fixed at the solar value \citepads[relative to the solar abundances derived by][]{1989GeCoA..53..197A}. Interstellar medium absorption was accounted for with the Tuebingen-Boulder absorption model (\texttt{tbabs}), with abundances set to \texttt{wilm} \citepads{2000ApJ...542..914W}. The fitting was performed with the ISIS spectral analysis package \citepads{2000ASPC..216..591H}.

In general, all available spectra pertaining to a given region were fit simultaneously. However, spectra obtained from data which were extracted from less than 80\% of the area of the data with largest spatial extent were not included; this is to avoid combining spectra taken from disparate areas within one region (an issue which may especially affect the larger regions).
Moreover, in those regions which contain both XMM-{\em Newton} and {\em Chandra} data, the respective {\em EPIC} and {\em ACIS} spectra were fit with the same spectral model but separately as combined fits did not generally yield acceptable results. For the spectra included in the simultaneous fit, all model parameters were tied, except for independent overall normalizations which were left to vary freely in order to account for differing spatial coverage due to observation dependent instrumental features such as CCD edges, bad pixels and chip gaps.

Comparing the three montage sets, the quality of the spectral fits in terms of $\chi^{2}_{\nu}$ tends to be worse in the 35 and 96 region sets than in the 419 region set. This is to be expected, as the former consists of coarser and larger regions which are more likely to straddle two or more distinct observational fields-of-view, and which, moreover, contain a large number of counts. Therefore, any calibration uncertainties in telescope vignetting and effective area would be more apparent in the resulting fits. In addition, physical properties varying within large regions may also yield worse fits given the model used. Whereas a single temperature model suffices for the finer
patchwork, the larger regions may require more complex models. In fact, spectral fits of large regions improve significantly in terms of $\chi^{2}_{\nu}$ when, for example, an absorbed two-temperatures shocked plasma is applied. For those regions where the spectral fit yielded $\chi^{2}_{\nu} \gtrsim$ 2, we mitigated the effect of the calibration uncertainties by imposing a stricter minimum coverage fraction, thus restricting the number of simultaneously fit spectra to those of greatest area coverage. A subsequent fit on this reduced data set would in general yield $\chi^{2}_{\nu} \lesssim$ 2.

\section{Results}
\label{sec:results}

We perfomed the first complete X-ray analysis of Puppis~A. As an example of the power of the method we include here the complete list of fit parameters for the 35-regions case (Table~\ref{tab:online}) and show maps of the spatial distribution of N$_{H}$, $kT$, O abundance and model normalization (proportional to the plasma emission measure, $\int{n_{e} n_{H} dV}$, where $n_{e}$ and $n_{H}$ are the electron and hydrogen densities respectively). Global conclusions, however, are drawn from the precisely fit 419-regions set. Due to the difficulty that represents in practice the labelling of 96 and/or 419 individual regions on an image, the derived parameters are not presented here, but are available upon request to the authors. Here we present the general and more interesting results that confirm the accuracy of our method (see Figures \ref{fig:panel1}, \ref{fig:panel2} and \ref{fig:panel3}).

\begin{figure}[ht!]
\centering
\includegraphics[scale=0.3,trim={0cm 0cm 0cm 0cm},clip]{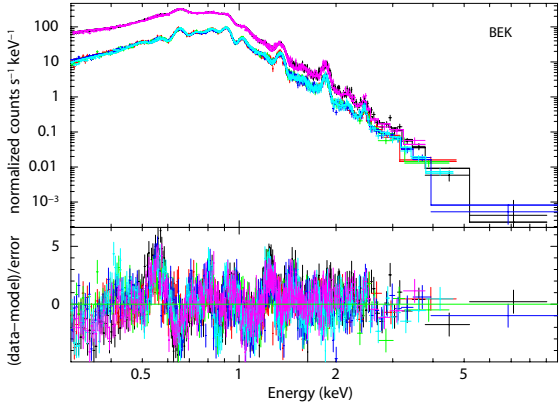}
\caption{Sample spectra ({\it color version available online}) from the region known as BEK, towards the eastern edge of Puppis~A, arising from the interaction of the SNR shock wave and a molecular cloud. Magenta and black data are from the XMM-{\em Newton} PN camera while blue, cyan, yellow, red, and green are data from the MOS camera. The resulting parameters from our fit of the whole region agree with average results obtained by \citetads{2005ApJ...635..355H} from a much detailed spatially-resolved study using {\it Chandra} observations.}
\label{fig:spec}%
\end{figure}

The method allows extraction of spectral information from the filamentary structures that are evident in the X-ray images. A similar analysis would otherwise require a careful manual definition and selection of those regions.  A sample spectrum, from the BEK, is displayed in Fig. \ref{fig:spec} together with the fit residuals in sigma units. The spectral model, an absorbed ($N_{H}=$ 0.38$_{-0.01}^{+0.01}$ $\times$10$^{22}$ cm$^{-2}$) single-temperature ($kT$=0.46$_{-0.01}^{+0.01}$ keV) plasma yields a moderate fit statistic ($\chi^{2}_{\nu}=1.85$/1989 d.o.f.). The fit improves significantly if we add a second plasma to the model ($\chi^{2}_{\nu}=1.36$/1965 d.o.f., $F$-test probability $< 1^{-10}$). This region has been thoroughly studied by \citetads{2005ApJ...635..355H} using observations obtained with {\em Chandra} and dividing the BEK into three main morphological components: a bright compact knot, a vertical curved structure (the bar) and a smaller bright cloud (the cap). On average, these regions were found to have temperatures in the 0.3--0.8~keV range and absorption columns in the 0.2--0.5$\times$10$^{22}$~cm$^{-2}$ range. Our method easily selects all these features as a single region (in the case of 35-regions) finding completely commensurate results.

The spectra from most regions have enough quality to allow us to constrain the elemental abundances. 
The spatial distribution of O, Ne, Mg, Si, and Fe is mostly smooth, with average and standard deviation (weighted by the count rate per region) across the remnant of O/O$_{\odot}$=0.20$\pm$0.01, Ne/Ne$_{\odot}$=0.39$\pm$0.03, Mg/Mg$_{\odot}$=0.27$\pm$0.01, Si/Si$_{\odot}$=0.37$\pm$0.07, S/S$_{\odot}$=0.78$\pm$2.28, and Fe/Fe$_{\odot}$=0.20$\pm$0.02. In agreement with the previous analysis by \citeads{2008ApJ...678..297K}, we found that the abundances are mostly subsolar. \citeads{2008ApJ...678..297K} tied the sulfur abundances to those of silicon, while we left both free to vary during the fit, which resulted in average sulfur abundance close to solar, although with too large dispersion (see Table \ref{tab:online}) to drawn meaningful conclusions.

 \begin{figure*}
   \centering
    \includegraphics[scale=0.8]{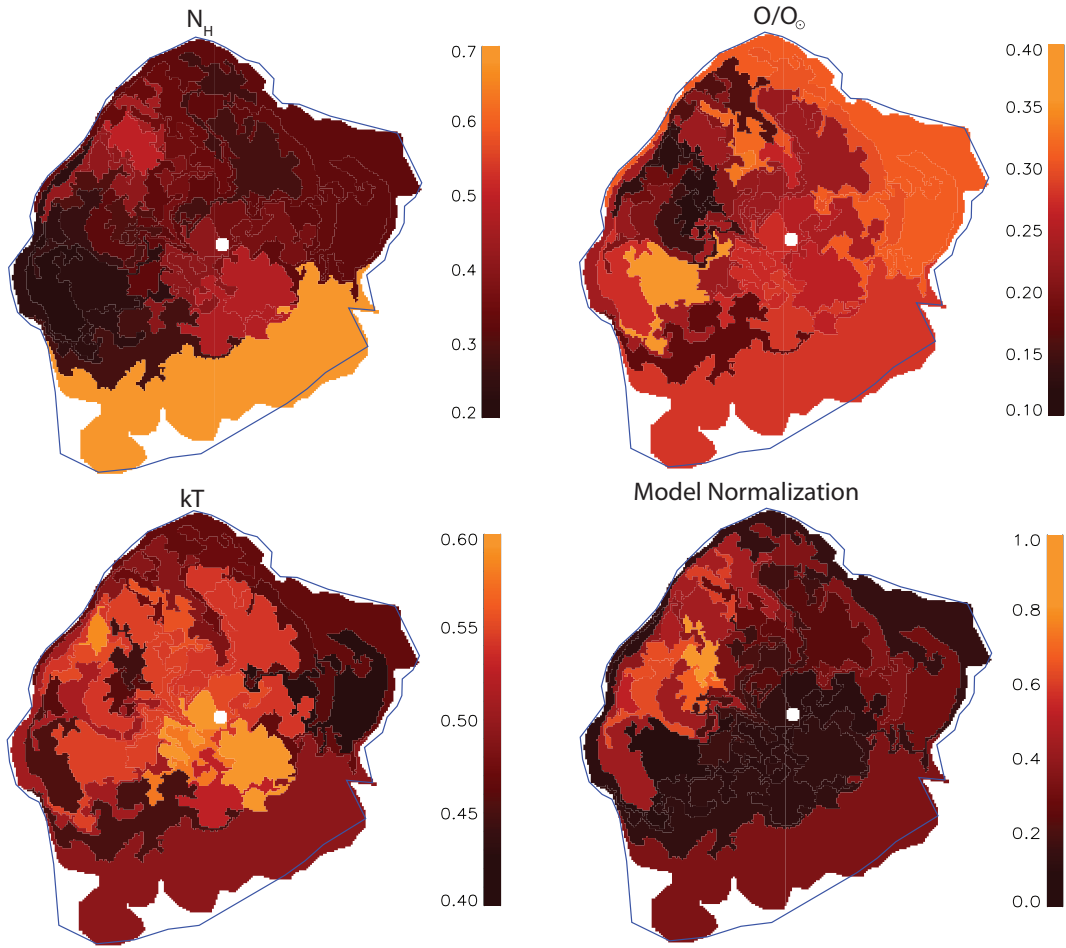}
   \caption{Sample color-coded maps ({\em color version available online}) of the spatial distribution of the parameters from the spectral fits. From left to right, this figure shows the spatial distribution of the shock temperature, $kT$, N$_{H}$, O/O$_{\odot}$ abundance and \texttt{vpshock} normalization (proportional to the plasma emission measure, $\int{n_{e} n_{H} dV}$, where $n_{e}$ and $n_{H}$ are the electron and hydrogen densities respectively).  }
    \label{fig:par}%
    \end{figure*}

\begin{figure}
\centering
\includegraphics[scale=0.7]{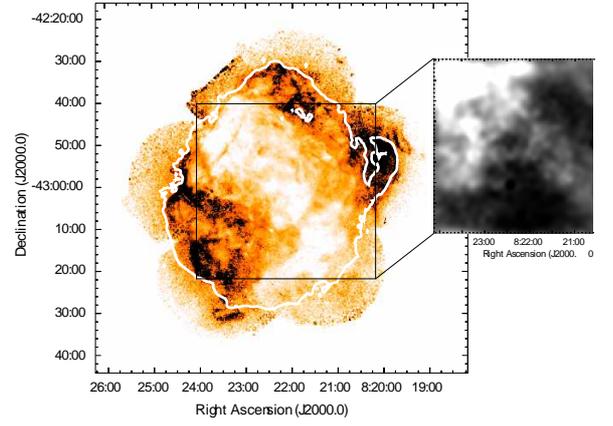}
\caption{Hardness ratio (0.3-0.7/1.0-8.0 keV) image ({\em color version available online}). The clear stripe crossing the remnant from East-North to South-West follows those regions where the hardness ratio is higher and so is the intervening absorption column. White contours show the continuum at radio wavelengths which delineate the SNR shock front. The inset shows distribution of N$_{H}$ integrated between 0 and 16.1 km s$^{-1}$ which is the systemic velocity of Puppis~A. The HI data were obtained by \citepads{2003MNRAS.345..671R} (compare with Fig. 3 in D13). }
\label{fig:hardness}%
\end{figure}

As shown in Fig. \ref{fig:panel3}, in the 35-regions case, the weak and diffuse X-ray emission toward the recently observed South and South-West portions of Puppis~A was selected by our algorithm as a single region (region \# 34). The spectrum from this region is well-described by a single-temperature shocked plasma with $kT$=0.49$_{-0.01}^{+0.01}$ keV, and abundances of O=0.28$_{-0.02}^{+0.02}$, Ne=0.49$_{-0.01}^{+0.01}$, Mg=0.20$_{-0.01}^{+0.02}$, Si=0.37$_{-0.03}^{+0.03}$, S=2.24$_{-0.37}^{+0.33}$ and Fe=0.20$_{-0.01}^{+0.01}$. Interestingly, we have also found an enhancement of absorption towards this region (see Fig.~\ref{fig:par}) with a value of $N_{H}$=0.68$_{-0.08}^{+0.01}\times$10$^{22}$ cm$^{-2}$ (the average $N_{H}$ across the whole remnant is 0.31$\times$10$^{22}$ cm$^{-2}$).  Fig.~\ref{fig:hardness} shows a comparison of the hardness ratio (0.3-0.7)/(1.0-8.0 keV) with the distribution of the absorbing column of atomic gas (as taken from D13), where it can be easily appreciated that the regions where the hardness ratio is higher have a good spatial concordance with the zones having higher N$_{H}$. As $kT$ tends to be larger towards the center of the SNR, temperature effects can also influence the hardness in the inner portions of the remnant.

\begin{figure}
\centering
\includegraphics[scale=0.5]{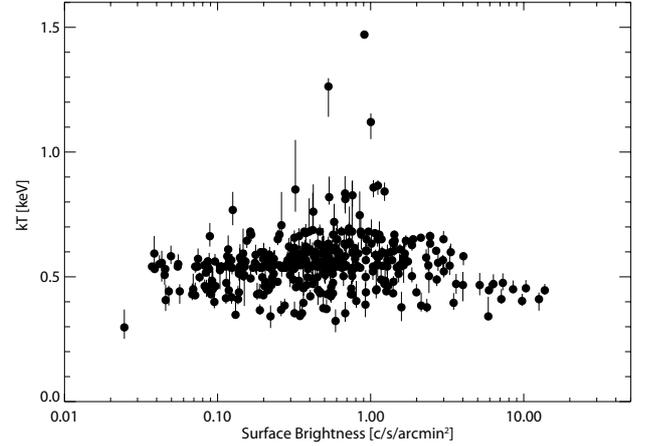}
\caption{Shock temperature versus surface brightness in the 419-regions case. While surface brightness varies over three orders of magnitudes throughout the remnant, the temperature changes by less than a factor of three.}
\label{fig:7}%
\end{figure}

The broadband X-ray image from Puppis~A reveals a very structured SNR with short curved filaments, suggesting that it is evolving in a complex environment. It is interesting to note that despite the intricate brightness morphology, the spatial distribution of plasma temperature, abundances and absorption is smooth. The plot of region brightness versus region temperature (Fig. \ref{fig:7}) confirms that although a wide range of brightness is present in Puppis~A, according with its filamentary nature, there is no significant temperature changes across the remnant. This finding suggests that the curved structures must be shaped by environmental inhomogeneities encountered by the shock that do not strongly affect the plasma temperature.

\section{Conclusions}
\label{sec:concl}

Our distinct technique is designed to extract spectral information from filamentary structures in extended X-ray sources, selecting them by their surface brightness, especially adequate for cases where different instruments are combined. As the spatial resolution can be tuned by using different bin sizes, the spectral analysis will accordingly reach that level of detail.  A big, middle-age, rich in structure SNR, which has been observed with different instrumental configurations throughout years, such as Puppis~A, represents an excellent pilot case to test this innovative technique to analyze the X-ray spectra along interesting features making it an excellent tool for X-ray studies of extended sources. Future work will include the application of  this ``feature-tailored'' selection of spectral regions to bright, young, ejecta-dominated SNRs. 

\begin{acknowledgements}
We acknowledge the anonymous referee for the useful comments that helped to improve the quality of this manuscript. GD, EG and GC acknowledge support from Argentina grants ANPCYT-PICT 2013-0902, ANPCYT-PICT 0571/11 and CONICET (Argentina) grant PIP 0736/11. GL, GD, EG and GC are members of the ``Carrera del Investigador Cient\'\i fico (CIC)" of CONICET.
\end{acknowledgements}

\bibliographystyle{aa}    
\bibliography{listaref_snr}
 
\onecolumn
\renewcommand{\arraystretch}{1.2}
\begin{landscape}
\begin{longtable}{ccccccccccccc}
\caption{Resulting parameters from the fit of an absorbed, variable abundances, plane-parallel shock model to the 35 regions selected with our feature-tailored technique with a count threshold per region of 2$\times$10$^{6}$ counts. Similar results with other threshold will be distributed upon request to the authors.}\\
\hline
\hline
Region \# & N$_{H}$[10$^{22}$ cm$^{-2}$] & kT [keV]  & O/O$_{\odot}$&  Ne/Ne$_{\odot}$ &  Mg/Mg$_{\odot}$ &  Si/Si$_{\odot}$ & S/S$_{\odot}$ & Fe/Fe$_{\odot}$ & $\tau$ [10$^{11}$ s cm$^{-3}$]  & $\chi^{2}_{\nu}$ & d.o.f  \\
\hline
\endfirsthead
\caption{Continued.} \\
\hline
Region \# & N$_{H}$[10$^{22}$ cm$^{-2}$] & kT [keV]  & O     &  Ne    &  Mg    &  Si     & S       &  Fe    & $\tau$ [10$^{11}$ s cm$^{-3}$]  & $\chi^{2}_{\nu}$ & d.o.f  \\
\hline
\endhead
\hline
\endfoot
\hline
\endlastfoot
0 & 0.229$_{-0.005}^{+0.008}$ & 0.454$_{-0.011}^{+0.003}$&  0.224$_{-0.010}^{+0.007}$& 0.437$_{-0.017}^{+0.015}$& 0.309$_{-0.010}^{+0.010}$& 0.427$_{-0.020}^{+0.018}$ & 0.894$_{-0.087}^{+0.109}$& 0.275$_{-0.012}^{+0.005}$& 3.304$_{-0.112}^{+0.119}$& 1.856& 1286 \\
1& 0.301$_{-0.007}^{+0.001}$ & 0.551$_{-0.005}^{+0.018}$&  0.187$_{-0.002}^{+0.002}$& 0.354$_{-0.004}^{+0.011}$& 0.254$_{-0.002}^{+0.003}$& 0.341$_{-0.018}^{+0.012}$& 0.686$_{-0.065}^{+0.054}$& 0.176$_{-0.004}^{+0.005}$ & 2.941$_{-0.033}^{+0.108}$& 1.510 & 1839 \\
2& 0.256$_{-0.006}^{+0.006}$& 0.512$_{-0.016}^{+0.005}$ &  0.203$_{-0.007}^{+0.009}$& 0.385$_{-0.013}^{+0.013}$& 0.279$_{-0.010}^{+0.013}$& 0.379$_{-0.013}^{+0.020}$& 0.696$_{-0.070}^{+0.079}$& 0.229$_{-0.012}^{+0.009}$& 2.328$_{-0.077}^{+0.041}$& 1.922    & 1331 \\
3& 0.277$_{-0.005}^{+0.004}$& 0.471$_{-0.009}^{+0.012}$& 0.303$_{-0.005}^{+0.006}$& 0.584$_{-0.010}^{+0.018}$& 0.434$_{-0.008}^{+0.009}$& 0.474$_{-0.032}^{+0.020}$& 1.015$_{-0.143}^{+0.117}$& 0.354$_{-0.003}^{+0.003}$& 1.860$_{-0.070}^{+0.059}$& 2.498& 706  \\
4& 0.402$_{-0.023}^{+0.008}$& 0.616$_{-0.028}^{+0.022}$& 0.223$_{-0.007}^{+0.012}$& 0.493$_{-0.002}^{+0.009}$ & 0.309$_{-0.009}^{+0.014}$& 0.340$_{-0.031}^{+0.027}$& 0.587$_{-0.111}^{+0.102}$ & 0.194$_{-0.004}^{+0.001}$& 4.160$_{-0.404}^{+0.492}$& 1.109    & 885  \\

5 & 0.382$_{-0.004}^{+0.001}$ & 0.543$_{-0.002}^{+0.004}$  &  0.332$_{-0.010}^{+0.009}$ & 0.653$_{-0.007}^{+0.012}$  & 0.567$_{-0.011}^{+0.012}$   & 0.758$_{-0.020}^{+0.025}$  & 1.498$_{-0.011}^{+0.012}$  & 0.181$_{-0.005}^{+0.005}$ &  1.367$_{-0.020}^{+0.030}$ & 2.477   & 2477 \\

6 & 0.220$_{-0.012}^{+0.003}$&0.454$_{-0.007}^{+0.023}$& 0.275$_{-0.006}^{+0.023}$  & 0.515$_{-0.011}^{+0.032}$& 0.384$_{-0.011}^{+0.011}$ & 0.493$_{-0.043}^{+0.029}$& 1.101$_{-0.202}^{+0.167}$& 0.329$_{-0.009}^{+0.032}$ & 2.366$_{-0.056}^{+0.111}$& 1.381& 1099 \\

7 & 0.273$_{-0.007}^{+0.002}$& 0.556$_{-0.010}^{+0.029}$& 0.302$_{-0.002}^{+0.020}$& 0.537$_{-0.004}^{+0.042}$& 0.387$_{-0.023}^{+0.026}$& 0.438$_{-0.046}^{+0.029}$ & 0.693$_{-0.151}^{+0.124}$& 0.268$_{-0.008}^{+0.018}$ & 2.403$_{-0.049}^{+0.137}$& 1.328& 1031 \\

8& 0.328$_{-0.019}^{+0.017}$& 0.561$_{-0.035}^{+0.032}$& 0.214$_{-0.009}^{+0.017}$& 0.420$_{-0.037}^{+0.041}$& 0.274$_{-0.030}^{+0.026}$ & 0.354$_{-0.038}^{+0.027}$& 1.001$_{-0.256}^{+0.241}$& 0.194$_{-0.013}^{+0.018}$ & 2.133$_{-0.091}^{+0.103}$& 1.268    & 846  \\

9 & 0.368$_{-0.003}^{+0.001}$& 0.541$_{-0.015}^{+0.008}$& 0.325$_{-0.001}^{+0.014}$& 0.595$_{-0.027}^{+0.008}$& 0.468$_{-0.001}^{+0.013}$ & 0.662$_{-0.054}^{+0.026}$& 1.172$_{-0.176}^{+0.132}$ & 0.313$_{-0.004}^{+0.007}$ & 1.893$_{-0.092}^{+0.014}$& 1.451  & 684  \\
10& 0.327$_{-0.013}^{+0.004}$ & 0.588$_{-0.034}^{+0.030}$& 0.252$_{-0.007}^{+0.040}$& 0.504$_{-0.043}^{+0.051}$& 0.348$_{-0.043}^{+0.038}$& 0.463$_{-0.057}^{+0.050}$& 0.994$_{-0.214}^{+0.215}$& 0.244$_{-0.024}^{+0.032}$& 2.566$_{-0.147}^{+0.095}$& 1.085    & 770  \\
11& 0.381$_{-0.027}^{+0.012}$ & 0.567$_{-0.040}^{+0.074}$  & 0.312$_{-0.033}^{+0.042}$& 0.653$_{-0.057}^{+0.088}$& 0.568$_{-0.054}^{+0.068}$& 1.076$_{-0.151}^{+0.102}$ & 2.077$_{-0.469}^{+0.493}$& 0.324$_{-0.052}^{+0.081}$& 1.625$_{-0.147}^{+0.299}$& 1.260    & 281  \\
12& 0.353$_{-0.003}^{+0.003}$ & 0.556$_{-0.010}^{+0.027}$&  0.320$_{-0.009}^{+0.004}$& 0.579$_{-0.019}^{+0.007}$& 0.441$_{-0.009}^{+0.010}$& 0.737$_{-0.053}^{+0.018}$ & 1.199$_{-0.156}^{+0.109}$& 0.296$_{-0.002}^{+0.003}$& 1.906$_{-0.042}^{+0.001}$& 1.583    & 717  \\

13& 0.292$_{-0.002}^{+0.001}$& 0.503$_{-0.008}^{+0.009}$ & 0.237$_{-0.005}^{+0.009}$ & 0.472$_{-0.002}^{+0.003}$ & 0.282$_{-0.006}^{+0.006}$ & 0.329$_{-0.015}^{+0.015}$ & 0.671$_{-0.060}^{+0.050}$ & 0.179$_{-0.001}^{+0.003}$& 2.644$_{-0.001}^{+0.005}$ & 1.649 & 2522 \\

14& 0.281$_{-0.009}^{+0.012}$& 0.544$_{-0.027}^{+0.015}$ & 0.274$_{-0.023}^{+0.021}$& 0.539$_{-0.034}^{+0.030}$& 0.329$_{-0.025}^{+0.022}$ & 0.370$_{-0.029}^{+0.035}$& 0.610$_{-0.142}^{+0.146}$& 0.228$_{-0.023}^{+0.018}$& 2.147$_{-0.065}^{+0.059}$& 1.380    & 1029 \\

15& 0.343$_{-0.003}^{+0.003}$& 0.484$_{-0.010}^{+0.020}$& 0.264$_{-0.002}^{+0.002}$ & 0.501$_{-0.002}^{+0.004}$ & 0.32$_{-0.01}^{+0.01}$ & 0.429$_{-0.040}^{+0.030}$& 0.715$_{-0.120}^{+0.100}$& 0.237$_{-0.003}^{+0.003}$ & 2.957$_{-0.080}^{+0.005}$  & 2.26    & 395 \\

16& 0.331$_{-0.006}^{+0.006}$& 0.459$_{-0.013}^{+0.014}$& 0.300$_{-0.007}^{+0.007}$& 0.548$_{-0.010}^{+0.008}$& 0.355$_{-0.007}^{+0.008}$& 0.461$_{-0.036}^{+0.023}$& 1.309$_{-0.234}^{+0.175}$ & 0.285$_{-0.010}^{+0.009}$& 1.487$_{-0.044}^{+0.109}$& 1.673    & 706  \\
17& 0.295$_{-0.010}^{+0.004}$& 0.536$_{-0.010}^{+0.011}$& 0.258$_{-0.008}^{+0.010}$ & 0.510$_{-0.006}^{+0.017}$& 0.331$_{-0.019}^{+0.012}$& 0.410$_{-0.029}^{+0.020}$& 0.791$_{-0.116}^{+0.103}$& 0.270$_{-0.002}^{+0.003}$& 2.018$_{-0.086}^{+0.073}$& 1.523    & 693  \\

18& 0.329$_{-0.001}^{+0.001}$& 0.542$_{-0.001}^{+0.001}$& 0.208$_{-0.001}^{+0.001}$ & 0.434$_{-0.001}^{+0.002}$& 0.285$_{-0.001}^{+0.004}$ & 0.386$_{-0.009}^{+0.012}$& 0.733$_{-0.090}^{+0.120}$  & 0.170$_{-0.002}^{+0.004}$ & 2.551$_{-0.001}^{+0.001}$ & 1.893   & 1622 \\

19& 0.232$_{-0.017}^{+0.003}$ & 0.546$_{-0.015}^{+0.047}$& 0.354$_{-0.007}^{+0.033}$&0.654$_{-0.008}^{+0.057}$& 0.373$_{-0.031}^{+0.039}$ & 0.447$_{-0.065}^{+0.054}$ & 0.845$_{-0.425}^{+0.394}$& 0.331$_{-0.013}^{+0.040}$ & 1.058$_{-0.067}^{+0.065}$& 1.332 & 861  \\
20 & 0.323$_{-0.008}^{+0.002}$& 0.426$_{-0.007}^{+0.014}$ & 0.311$_{-0.001}^{+0.001}$& 0.574$_{-0.003}^{+0.003}$& 0.291$_{-0.013}^{+0.008}$ & 0.472$_{-0.043}^{+0.033}$ & 3.005$_{-0.687}^{+0.440}$ & 0.258$_{-0.001}^{+0.002}$& 0.973$_{-0.022}^{+0.027}$& 1.561    & 1049 \\
21& 0.363$_{-0.001}^{+0.001}$& 0.546$_{-0.001}^{+0.001}$& 0.224$_{-0.001}^{+0.004}$ & 0.492$_{-0.001}^{+0.007}$& 0.323$_{-0.001}^{+0.006}$ & 0.427$_{-0.008}^{+0.006}$& 0.874$_{-0.030}^{+0.032}$ & 0.184$_{-0.001}^{+0.003}$& 1.921$_{-0.001}^{+0.027}$& 1.706    & 5954 \\

22 & 0.320$_{-0.001}^{+0.001}$& $\lesssim$0.543  &  $\lesssim$0.232 & $\lesssim$0.476  & $\lesssim$0.318  & 0.465$_{-0.009}^{+0.010}$ & 0.784$_{-0.050}^{+0.045}$  & $\lesssim$0.233 & $\gtrsim$1.880 & 2.576    & 1890 \\

23 & 0.236$_{-0.006}^{+-0.003}$& 0.507$_{-0.014}^{+0.018}$ & 0.233$_{-0.001}^{+0.002}$& 0.437$_{-0.003}^{+0.004}$& 0.312$_{-0.009}^{+0.007}$& 0.393$_{-0.028}^{+0.019}$& 1.053$_{-0.168}^{+0.133}$& 0.268$_{-0.005}^{+0.007}$& 1.528$_{-0.060}^{+0.096}$& 1.722    & 1526 \\
24& 0.321$_{-0.001}^{+0.001}$& 0.546$_{-0.001}^{+0.006}$& 0.226$_{-0.001}^{+0.004}$& 0.475$_{-0.001}^{+0.007}$& 0.266$_{-0.001}^{+0.005}$& 0.338$_{-0.008}^{+0.006}$& 0.789$_{-0.037}^{+0.034}$& 0.172$_{-0.001}^{+0.003}$& 1.717$_{-0.001}^{+0.001}$& 1.675    & 6146 \\

25& $\gtrsim$0.295 & $\lesssim$0.543  & $\lesssim$0.234 & $\lesssim$0.462 & $\lesssim$0.285  & 0.374$_{-0.009}^{+0.009}$& 0.751$_{-0.041}^{+0.047}$ & $\lesssim$0.238  & $\lesssim$1.726  & 2.859    & 1988 \\

26& 0.324$_{-0.016}^{+0.012}$& 0.467$_{-0.031}^{+0.051}$& 0.309$_{-0.019}^{+0.015}$ & 0.561$_{-0.040}^{+0.021}$& 0.307$_{-0.044}^{+0.026}$ & 0.536$_{-0.116}^{+0.115}$& \nodata & 0.288$_{-0.011}^{+0.018}$& 0.590$_{-0.024}^{+0.071}$& 1.085    & 624  \\

27& 0.364$_{-0.003}^{+0.003}$ & 0.551$_{-0.005}^{+0.005}$  & 0.253$_{-0.002}^{+0.002}$ & 0.487$_{-0.003}^{+0.003}$ & 0.286$_{-0.004}^{+0.006}$& 0.362$_{-0.010}^{+0.003}$ & 0.809$_{-0.059}^{+0.060}$ & 0.225$_{-0.002}^{+0.007}$& 1.772$_{-0.010}^{+0.012}$& 2.269& 1234 \\

28& $\lesssim$0.408 & 0.593$_{-0.010}^{+0.009}$  & 0.272$_{-0.002}^{+0.002}$ & 0.541$_{-0.003}^{+0.003}$& 0.325$_{-0.005}^{+0.005}$& 0.437$_{-0.020}^{+0.010}$  & 1.015$_{-0.007}^{+0.007}$ & 0.200$_{-0.003}^{+0.003}$& 1.329$_{-0.010}^{+0.031}$ & 2.762 & 891 \\

29& 0.402$_{-0.002}^{+0.002}$ & 0.577$_{-0.006}^{+0.009}$ & 0.272$_{-0.002}^{+0.005}$ & 0.571$_{-0.005}^{+0.007}$ & 0.302$_{-0.005}^{+0.005}$ & 0.363$_{-0.020}^{+0.017}$ & 1.071$_{-0.011}^{+0.012}$ & 0.194$_{-0.001}^{+0.002}$ & 1.157$_{-0.021}^{+0.023}$ & 2.160    & 814 \\

30& 0.368$_{-0.001}^{+0.001}$ & 0.541$_{-0.001}^{+0.003}$ & 0.242$_{-0.001}^{+0.001}$ & 0.498$_{-0.001}^{+0.009}$ & 0.270$_{-0.001}^{+0.002}$ & 0.350$_{-0.012}^{+0.014}$ & 1.022$_{-0.006}^{+0.007}$ & 0.221$_{-0.001}^{+0.006}$  & 1.247$_{-0.021}^{+0.007}$ & 2.105    & 1961 \\

31& 0.472$_{-0.005}^{+0.002}$ & 0.526$_{-0.016}^{+0.022}$& 0.282$_{-0.010}^{+0.012}$ & 0.539$_{-0.012}^{+0.020}$ & 0.221$_{-0.010}^{+0.012}$ & 0.310$_{-0.010}^{+0.013}$ & 0.933$_{-0.200}^{+0.300}$ & 0.223$_{-0.003}^{+0.007}$ & 1.149$_{-0.049}^{+0.025}$ & 2.230    & 403 \\

32& 0.281$_{-0.008}^{+0.011}$& 0.450$_{-0.020}^{+0.012}$& 0.174$_{-0.012}^{+0.006}$& 0.341$_{-0.023}^{+0.013}$& 0.245$_{-0.017}^{+0.011}$& 0.361$_{-0.030}^{+0.024}$& 1.597$_{-0.372}^{+0.420}$& 0.207$_{-0.016}^{+0.008}$& 1.267$_{-0.037}^{+0.048}$& 1.494    & 799  \\

33& 0.476$_{-0.001}^{+0.001}$& 0.593$_{-0.001}^{+0.014}$& 0.259$_{-0.001}^{+0.004}$ & 0.521$_{-0.001}^{+0.003}$  & 0.236$_{-0.001}^{+0.002}$  & 0.305$_{-0.010}^{+0.011}$  & 0.833$_{-0.070}^{+0.080}$ & 0.203$_{-0.001}^{+0.002}$ & 1.045$_{-0.020}^{+0.024}$ & 2.921    & 1960 \\

34& 0.678$_{-0.008}^{+0.005}$& 0.494$_{-0.013}^{+0.013}$& 0.281$_{-0.015}^{+0.016}$& 0.496$_{-0.004}^{+0.006}$& 0.211$_{-0.009}^{+0.006}$& 0.373$_{-0.023}^{+0.017}$ & 2.242$_{-0.372}^{+0.327}$& 0.202$_{-0.002}^{+0.002}$& 0.763$_{-0.016}^{+0.035}$& 1.640    & 1462

\label{tab:online}
\end{longtable}

\end{landscape}
\twocolumn

\end{document}